\documentclass[journal]{IEEEtran}

\ifCLASSINFOpdf
\else
   \usepackage[dvips]{graphicx}
\fi
\usepackage{url}

\usepackage{amssymb}
\usepackage{amsmath}
\usepackage{mathrsfs}
\usepackage{bm}
\usepackage{array}
\usepackage{graphicx}
\usepackage{orcidlink}
\usepackage{tikz}
\usepackage{multirow}
\usepackage[numbers,sort&compress]{natbib}
\usepackage[ruled,linesnumbered]{algorithm2e}
\usepackage{bold-extra}
\usepackage{booktabs}
\usepackage{float}
\usepackage{makecell}
\usepackage{threeparttable}
\usepackage{subfigure}
\usepackage[subfigure]{tocloft}
\usepackage{xcolor} 
\usepackage{soul} 
\usepackage{balance}
\renewcommand{\arraystretch}{0.25}
\makeatletter

\newcommand{\Rmnum}[1]{\expandafter\@slowromancap\romannumeral #1@}
\sethlcolor{yellow}
\makeatother

\hypersetup{
colorlinks=true,
linkcolor=black,
citecolor=black
}

\begin{document}
\vspace{-\baselineskip}
\title{Deep Learning based Performance Testing for Analog Integrated Circuits}

\author{Jiawei Cao, 
Chongtao Guo,
Hao Li, 
Zhigang Wang, 
Houjun Wang, 
and Geoffrey Ye Li, \IEEEmembership{Fellow, IEEE}
\vspace{-5mm}
\thanks{This work was supported by National Natural Science Foundation of China (62303091), and supported by China Postdoctoral Science Foundation
(2021M700707), and supported by Natural Science Foundation of Sichuan
Province (2022NSFSC0905), and supported by Fundamental Research Fund
for the Central University of China (ZYGX2022J015).}
\thanks{(Corresponding author: Hao Li.)}
\thanks{J. Cao is with the School of Automation Engineering, University of Electronic Science and Technology of China, Chengdu 611731, China, and also with the Shenzhen Institute for Advanced Study, University of Electronic Science and Technology of China, Shenzhen 518110, Guangdong, China (e-mail: joe\_gavin@163.com).} 
\thanks{C. Guo is with the Guangdong Key Laboratory of Intelligent Information Processing, College of Electronics and Information Engineering, Shenzhen University, Shenzhen 518060, China (e-mail: ctguo@szu.edu.cn).}
\thanks{H. Li, Z. Wang are with the School of Automation Engineering, University of Electronic Science and Technology of China, Chengdu 611731, China (haoli@uestc.edu.cn; wangzhigang@uestc.edu.cn).}
\thanks{H. Wang is with the Shenzhen Institute for Advanced Study, University of Electronic Science and Technology of China, Shenzhen 518110, Guangdong, China (e-mail: hjwang@uestc.edu.cn).}
\thanks{Geoffrey Ye Li is with the Department of Electrical and Electronic Engineering, Imperial College London, SW7 2AZ London, U.K. (e-mail:
geoffrey.li@imperial.ac.uk).}}

\maketitle
\begin{abstract}
In this paper, we propose a deep learning based performance testing framework to minimize the number of required test modules while guaranteeing the accuracy requirement, where a test module corresponds to a combination of one circuit and one stimulus.
First, we apply a deep neural network (DNN) to establish the mapping from the response of the circuit under test (CUT) in each module to all specifications to be tested.
Then, the required test modules are selected by solving a 0-1 integer programming problem.
Finally, the predictions from the selected test modules are combined by a DNN to form the specification estimations.
The simulation results validate the proposed approach in terms of testing accuracy and cost.
\end{abstract}

\IEEEpeerreviewmaketitle
\section{Introduction}
\IEEEPARstart
{P}{ost}-package analog integrated circuit (IC) testing aims to measure the specifications that characterize its performance.
With carefully selected stimulus signals imposed on specifically designed test circuits, traditional methods measure ICs by analyzing the corresponding response signals.
Almost every specification requires a particular test module to extract the corresponding feature, where a test module corresponds to a specific combination of a stimulus and a circuit. 
Therefore, the test costs are usually high from the perspectives of time consumption and hardware resources, especially when the number of specifications in huge in modern advanced ICs. \par
To reduce the test costs, a key idea is to minimize the number of required test modules.
The work in \cite{milor_optimal_1990} has analyzed the relation between fault coverage (FC) and test modules, which has been further leveraged to optimize the testing order.
Following this order, the rest of the testing task can be skipped once a fault is detected.
Therefore, addressing the testing order arrangement problem, with an original complexity $\mathcal{O}(n!)$, plays the most prominent role for such an approach.
To this end, the dynamic programming proposed in \cite{huss_optimal_1991} has cut down the complexity of the testing order optimization to $\mathcal{O}(n^22^n)$, which makes such kind of cost reducing method to be more popular. 
Although the testing order arrangement based approach can be well applied to detect IC faults, it is helpless for performance testing where all specifications need to be measured in terms of their exact values. \par
Machine learning (ML) can characterize the mapping from a test module's response to multiple specifications \cite{variyam_prediction_2002}, which could fully exploit the capability of one module in testing multiple specifications and further reduce the number of the required modules.
Leveraging ML technology, intelligent performance testing strategies need fewer modules to satisfy the testing accuracy requirement. 
Towards this direction, one of the most fundamental issues is to explore the potential of a single module in predicting multiple specifications, which has been investigated from various perspectives, including hardware, data, and algorithms. 
First, on the hardware level, a nonlinear defect filter for analog ICs has been proposed in \cite{stratigopoulos_filter_2009}, by which the defective samples are filtered out to ensure a reliable set of circuit under test (CUT) samples are used in the training stage.
The filter has been applied in \cite{stratigopoulos_adaptive_2018} and \cite{badawi2021evaluation} to screen out the samples with suspicious performance in the test stage.
Moreover, the genetic algorithm (GA) has been adopted in \cite{deyati_dynamic_2020} and \cite{banerjee_automatic_2015} to generate the optimal stimulus for intelligent testing.
Second, on the data processing aspect, the researchers in \cite{barragan_use_2019} and \cite{leger_brownian_2016} have committed to finding the connection between the tested performance space and input space to determine the most effective pre-processing. 
Besides the responses, the research in \cite{hou2022statistical} has utilized a part of specifications that are easy to test as the inputs to estimate the other specifications that are difficult to test. 
Third, in terms of training algorithms, multivariate adaptive spline regression (MARS) in \cite{variyam_prediction_2002}, \cite{deyati_dynamic_2020} and neural networks in \cite{stratigopoulos_filter_2009}, \cite{hou2022statistical}, \cite{oneshot2016} are the two most widely utilized training approaches for their good non-linear regression ability.\par
With a single module taken into consideration, any particular one may have the inadequate ability to test some of the specifications to meet the accuracy requirements. Hence, it is a challenge for the aforementioned methods to test all specifications and satisfy corresponding requirements. 
In this paper, we initially allow multiple test modules to coexist and adopt a deep neural network (DNN) for each test module to establish a mapping from the test modules' responses to the specifications of CUTs.
Then, we select the required modules by addressing a 0-1 integer programming problem to reduce the test cost.
Finally, we train an additional DNN to combine the selected predictions to improve test accuracy. \par
\color{black}
\section{System Model}
The aim of testing performance of analog ICs is achieved by analyzing the responses triggered by a number of stimuli over a number of test circuits. 
We consider $N$ available time-domain stimuli and the $n$th stimulus given by $s_n(t)$.
There are $M$ test circuits, where the $m$th circuit's system function mapping the input signal to the output signal is denoted by $h_m(\cdot)$.
Particularly, the response of $s_n(t)$ over the $m$th test circuit is denoted by
\begin{equation}
   r_{m,n}(t)= h_m(s_n(t)).
   \label{R-t}
\end{equation}
Different combinations of stimuli and test circuits constitute differences test modules, leading to the total number of test modules being $M\!N$.\par

Consider a CUT with $L$ key specifications, where the ground-truth value of the $\ell$th one is denoted by $p(\ell)$, which form a vector $\mathbf{p}= [p(1),p(2),...,p(L)]^\intercal \in \mathbb{R}^{L\times 1}$.
The CUT testing here is to derive an estimation of $\mathbf{p}$, denoted by $\hat{\mathbf{p}}$, by exploiting the response signals in different test modules.
In particular, from the module with the $m$th circuit stimulated by the signal $s_n(t)$, we have the estimation, denoted by $\hat{\mathbf{p}}_{m,n}$, given by
\begin{equation}
   \hat{\mathbf{p}}_{m,n}= f_{m,n}(r_{m,n}(t)) \in \mathbb{R}^{L\times 1},
   \label{mapping}
\end{equation}
where $f_{m,n}(\cdot)$ is the mapping function to be determined.
In this work, we adopt a DNN with parameters ${\pmb{\phi}}_{m,n}$ to characterize mapping $f_{m,n}(\cdot)$, where ${\pmb{\phi}}_{m,n}$ are trained by a sufficient number of CUTs with their ground-truth specification labels.
However, each test module may have a distinct observation capability for measuring different specifications.
In other words, a test module may be good at estimating one specification but help little in estimating another.
For instance, a DC stimulus signal is effective in testing DC specifications, such as offset voltage (VOS) for an amplifier, but is helpless for testing AC specifications like phase margin (PM).
To this end, exploiting the selected test modules collaboratively to satisfy the accuracy requirements of all specification and optimize the test costs is possible, and desired.

\section{Proposed Intelligent Testing Approach}

In this section, we will first focus on the training for all test modules and then we will investigate the module selection method and the strategy of combining the predictions of the selected modules. Finally, we will briefly introduce data processing in the testing stage.
\subsection{Module Training}
Consider that we have $W$ CUT samples for training, whose specification labels form a matrix $\mathbf{P} =[\mathbf{p}^1,\mathbf{p}^2,\cdots, \mathbf{p}^W]^\intercal \in \mathbb{R}^{W \times L}$ that can be derived by the traditional method, where $\mathbf{p}^w \in \mathbb{R}^{L\times 1}$ is the vector formed by the $w$th CUT's $L$ specification labels.
To impose the stimulus onto the input of DNN, we evenly sample the continuous-time responses of the $W$ CUTs with a period of $\tau$, leading to a $W$-by-$K$ matrix
\begin{equation}
   \mathbf{R}_{m,n}=
   \begin{bmatrix}
      r_{m,n}^1(\tau)& r_{m,n}^1(2\tau)& \cdots& r_{m,n}^1(K\tau) \\
      r_{m,n}^2(\tau)& r_{m,n}^2(2\tau)& \cdots& r_{m,n}^2(K\tau) \\
      \vdots&  \vdots& \ddots          & \vdots                   \\
      r_{m,n}^W(\tau)& r_{m,n}^W(2\tau)& \cdots& r_{m,n}^W(K\tau) \\
   \end{bmatrix},\\
   \label{R}
\end{equation}
where ${r}_{m,n}^w(k\tau)$ is the $k$th sample of the $w$th CUT's response signal. \par
The amplitude difference among the different CUTs' responses at every specific time is much smaller than the amplitude range of the response from time $\tau$ to time $K\tau$, which will result in a poor regression effect.
By normalizing the responses at each time point, i.e., normalizing each column of the matrix $\mathbf{R}_{m,n}$, we can obtain $\mathbf{\hat{R}}_{m,n}$.
Specifically, the element in the $w$th row and the $k$th column of $\mathbf{\hat{R}}_{m,n}$, i.e., $\hat{r}_{m,n}^w(k\tau)$, can be expressed as
\begin{equation}
   \hat{r}_{m,n}^w(k\tau)=\frac{r_{m,n}^w(k\tau)-\mu_{m,n}(k\tau) }{\sigma_{m,n}(k\tau)},
   \label{hat-r}
\end{equation}
where $\mu_{m,n}(k\tau)$ and $\sigma_{m,n}(k\tau)$ are respectively given by
\begin{equation}
   \mu_{m,n}(k\tau)=\frac{1}{W}\sum_{w=1}^Wr_{m,n}^w(k\tau)
   \label{mean}
\end{equation}
and
\begin{equation}
   \sigma_{m,n}(k\tau)=\sqrt{\frac{1}{W}\sum_{w=1}^W(r_{m,n}^w(k\tau)-\mu_{m,n}(k\tau))^2}.
   \label{var}
\end{equation}
Taking all rows of $\mathbf{\hat{R}}_{m,n}$ as inputs and all rows of $\mathbf{P}$ as the corresponding labels, we can train a DNN with parameters ${\pmb{\phi}}_{m,n}$ for the $m$th circuit and the $n$th stimulus.
For ${\pmb{\phi}}_{m,n}$, the loss function is expressed as
\begin{equation}
   \mathcal{L}_{m,n}=\frac{1}{W}\sum_{w=1}^W\Vert \mathbf{p}^w-\hat{\mathbf{p}}_{m,n}^w\Vert^2,
   \label{loss}
\end{equation}
where vector $\hat{\mathbf{p}}_{m,n}^w =[\hat{p}_{m,n}^w(1),\hat{p}_{m,n}^w(2),\ldots,\hat{p}_{m,n}^w(L) ]^\intercal$ denotes the prediction of $\mathbf{p}^w$, $\hat{p}_{m,n}^w(\ell)$ denotes the $\ell$th specification in $\hat{\mathbf{p}}_{m,n}^w$. 

\subsection{Module Selection}
Based on the premise that every tested specification satisfies the accuracy requirement, to reduce the test costs, we should select as few stimuli and test circuits as possible in testing.\par
Let $\mathbf{x}:=\{x_{m,n}|\forall m \in [M], \forall n \in [N]\} $ denote the set of decision variables,
where $[M] := \{1,2,\ldots,M\}$ denote the running index set induced by integer $M$, $x_{m,n}\!=\!1$ means that the test module consisting of the $m$th test circuit and the $n$th stimulus is selected, and $x_{m,n}\!=\!0$ otherwise.
The mean-squared error (MSE) between the model prediction value and the actual value is adopted to evaluate the test accuracy in this paper.
In particular, the $\ell$th specification's MSE predicted by $\pmb{\phi}_{m,n}$ is given by
\begin{equation}
   e_{m,n}{(\ell)} = \frac{1}{W}\sum_{w=1}^{W} \left(p^{w}(\ell)-\hat{p}_{m,n}^{w}(\ell) \right)^2.
\label{MSE}
\end{equation}
For each specification, to ensure that at least one of the selected test modules predicts this specification satisfied the corresponding accuracy requirement, we use the minimum MSE of the selected modules to evaluate its test performance. 
Let $\epsilon_\ell$ denote the MSE threshold of the $\ell$th specification.
Then, we can formulate the module selection (MS) problem as a 0-1 integer programming issue given by
\begin{equation}
   \begin{aligned}
      \min_{x_{m,n} \in \mathbf{x}}& \quad \sum_{m=1}^{M} \sum_{n=1}^{N} \lambda_{m,n}x_{m,n}\\
      \text{s.t.} &\quad \min \{e_{m,n}(\ell) \mid x_{m,n}=1\} \leq \epsilon_\ell, \forall \ell \in [L]\\
      &\quad \sum_{m=1}^{M} \sum_{n=1}^{N} x_{m,n} \geqslant 1\\
      &\quad x_{m,n} \in \{0, 1\}, \forall n \in [N], \forall m \in [M],
      \label{0-1}
      \end{aligned}
\end{equation}
where $\lambda_{m,n}$ is the module cost of the $m$th test circuit and the $n$th stimulus.
The optimal solution of Problem (9), denoted by $\mathbf{x}^*=\{x^*_{m,n}|\forall m \! \in \! [M], \forall n \! \in \! [N]\}$, indicates the selection situation of each test module when all specifications respectively reach the required accuracy. In this situation, and the test costs are minimized.
There are $2^{M\!N}$ feasible selection situations in Problem (\ref{0-1}) and the computational complexity will reach $\mathcal{O}(2^{M\!N}L)$ if exhaustive search method is adopted.
Thus, we turn to the implicit enumeration algorithm \cite{implicit}, which can solve the 0-1 integer programming with more efficiency.

\subsection{Results Combination and Testing}
It is essential to investigate the approaches to deal with different predictions because the solution of Problem (\ref{0-1}) may indicate that more than one test module need to be selected.
If the final test value of one specification is assigned as the prediction corresponding to the minimum, the information provided by other predictions will be wasted.
Weighted sum (WS) is a straightforward way to exploit the predicted values comprehensively, but such a linear combination model cannot capture the nonlinear relations that may exist between the final test specification and the corresponding selected predictions.
Therefore, a DNN with $TL$-dimensional inputs and $L$-dimensional outputs is adopted to combine the predictions since it has both linear and nonlinear regression ability, where $T=|\{x^*_{m,n}|x^*_{m,n}=1\}|$ is the number of selected modules. 
In detail, for each training sample, the predictions generated by the selected test modules are utilized as inputs, while the specifications obtained from traditional testing methods serve as labels, to train the parameters $\pmb{\rho}$ of the DNN.\par

In the test stage, we use the selected stimuli and test circuits to generate responses, which are pre-processed by (\ref{hat-r}). 
It is worth noting that the used $\mu_{m,n}(k\tau)$ and $\sigma_{m,n}(k\tau)$ are those calculated and saved during the training stage. \par
\section{Experiment results and analysis}
\begin{figure}[t]
   \centering
   \includegraphics[scale=0.4]{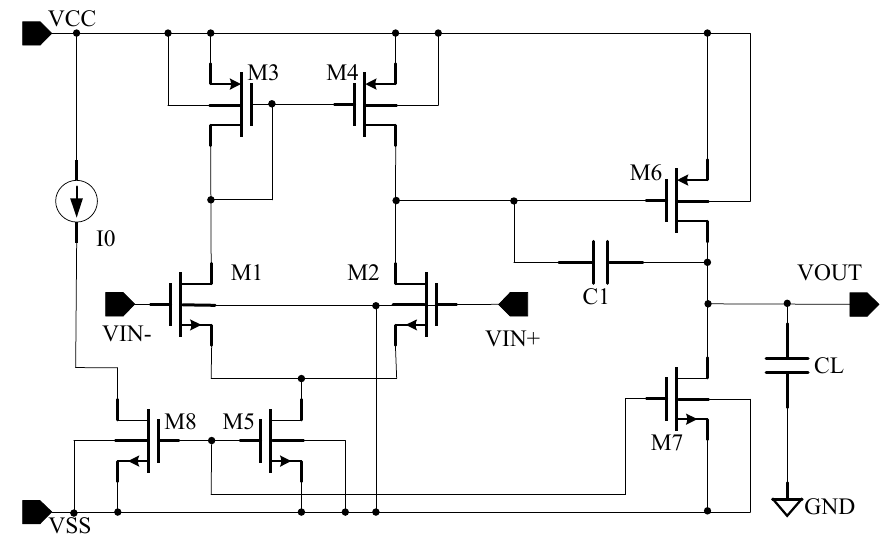}
   \vspace{-0.5\baselineskip}
   \caption{Schematic diagram of OPAMP in simulation.}
   \label{OPA}
\end{figure}

In this section, we evaluate the testing performance of the proposed framework by simulation experiments. 
We develop a operational amplifier (OPAMP) as the CUT, where the circuit is designed with TSMC-180 nm process design kits (PDK).
The schematic diagram of the OPAMP is shown in Fig. \ref{OPA}. 
Ten representative specifications in typical application situations are chosen as test objectives.
Generally, determining the optional test modules all types of specifications, i.e., AC, DC, and transient specifications, should be considered.
Some special situations, such as CUT working in a nonlinear scenarios, should be take into account, either.
For this reason, the module training involves four stimuli and two test circuits, where
the stimuli include chirp signal, random signal, two-tone signal, and pulse.
The two test circuits have amplification factors of three and ten times, which can be realized by different feedback resistors.\par
Total 5,000 instances of the CUT are generated by the Monte-Carlo simulation in Spectre, where the instance ratio of training to testing is 7:3. 
The label dataset is established by eight test modules with the traditional method.
The cost of each test module $\lambda_{m,n}$ is a personalized value based on the practical test environment, which can be determined by test engineers. 
The value of $\lambda_{m,n}$ will affect the solutions of Problem (\ref{0-1}), rather than its modeling process.
In the simulation environment, we set all $\lambda_{m,n}$ values to 1.
The primary information of the CUT and key parameters of DNNs are listed in the upper and lower part of TABLE \ref{setup-list}, respectively.
\par
\begin{figure*}[h]
   \vspace{-\baselineskip}
   \centering
   \includegraphics[width=0.85\linewidth]{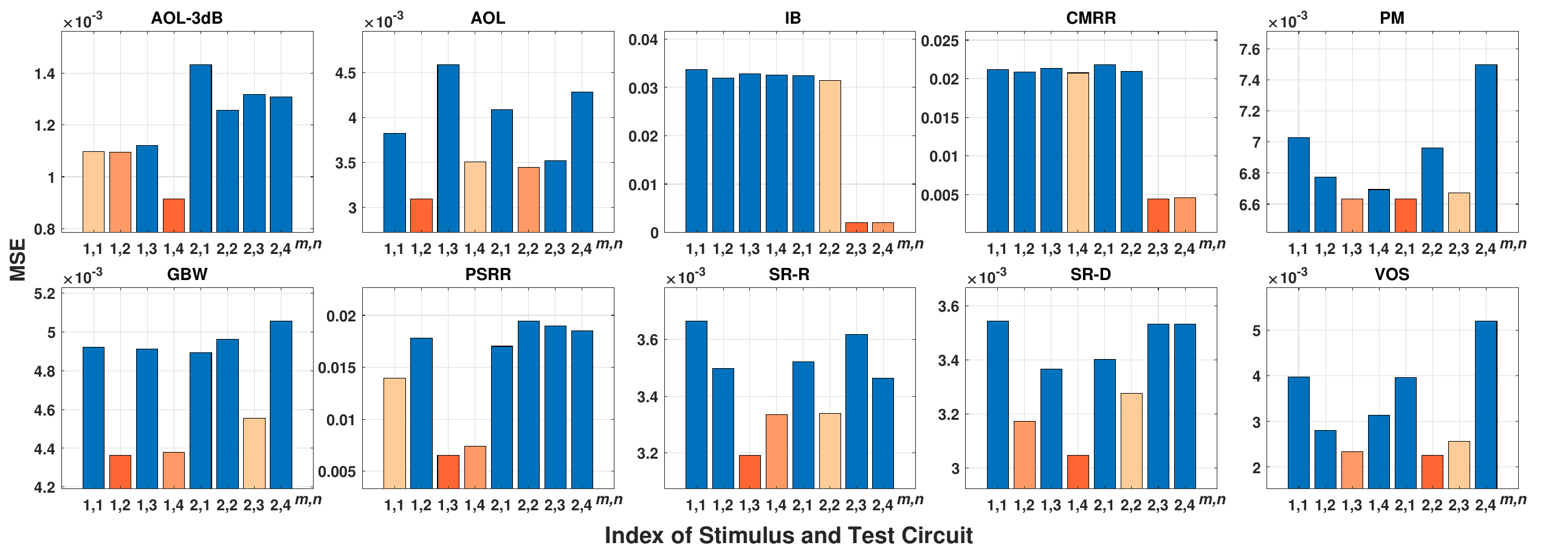}
   \vspace{-\baselineskip}
   \caption{MSE of prediction in all test modules, where the the test module corresponding to the $m$th circuit and the $n$th stimulus, denoted by $m$-$n$.}
   \label{mse-10}
\end{figure*}
\begin{table}[t]
   \vspace{-1.5\baselineskip}
   \renewcommand{\arraystretch}{1}
   \centering
   \caption{Experimental Setup}
   \label{setup-list}
   \begin{threeparttable}
       \begin{tabular} 
         {p{3cm} p{5cm}}
         \toprule[1pt]
         \midrule
         \textbf{ITEM} &  \textbf{CONDITION} \\ 
         \midrule
         \textbf{CUT} & OPAMP \\
         \textbf{Process} & TSMC-180 nm \\
         \textbf{Samples} & Number:5000 $|$ Ratio:7:3 (training:testing)\\
         \textbf{Simulator} & Spectre\\
         \textbf{Stimuli} & 4 (random $|$ chirp $|$ pulse $|$ double-tone)\\
         \textbf{Responses} & Time: 10 us $|$ Points: 10001\\
         \textbf{Test Circuits} & 2 (negative feedback: $\times 3$ $|$ $\times 10$)\\
         \textbf{Specifications} & 10 (AC:7 $|$ DC:1 $|$ Transient:2)\\
         \textbf{Module Cost} & $\lambda_{m,n}=1, \forall m,n$\\
         \textbf{MSE threshold} & AOL-3dB:1.1 $|$ AOL:4.2 $|$ IB:9.7\\
         \textbf{(unit:$\times 10^{-3}$)} & CMRR:7.5 $|$ PM:6.9 $|$ GBW:4.8 $|$ PSRR:1.1\\
         \textbf{} &SR-R:3.7 $|$ SR-D:3.6 $|$ VOS:2.7\\
         \midrule
         \textbf{Net Type} & DNN\\
         \textbf{Activation Function} & ReLU\\
         \textbf{Architecture} & fully-connected\\
         \textbf{Loss Function} & MSE \\
         \textbf{Optimizer} & Adam \\
         \textbf{Learning Rate} & $5 \times 10^{-4}$ \\
         \textbf{Number of Layers} & ${\pmb{\phi}}_{m,n}$:7\ \ \ \ \ \  $\pmb{\rho}$:5\\
         \textbf{Batch Size} & $\pmb{\phi}_{m,n}$:32 \ \ \ \ \!$\pmb{\rho}$:16\\
         \textbf{Epchos} & $\pmb{\phi}_{m,n}$:100 \ \ $\pmb{\rho}$:75\\
       \bottomrule[1pt]
       \end{tabular}
       \vspace{-\baselineskip}
   \end{threeparttable}
\end{table}

The MSE of every specification predicted by DNNs is shown in Fig. \ref{mse-10}.
The predictions of each specification are placed in a sub-figure, where we mark the three lowest MSE in warm colors.
The results confirm the diverse capabilities of every test module to predict different specifications, e.g.,the test module corresponding to the second circuit and the fourth stimulus, denoted by 2,4, obtains the lowest MSE when predicting IB while the highest MSE when predicting PM, PSRR, and VOS.
It is thus not straightforward to select a minimum number of test modules to meet the MSE requirements of all specification predictions, necessitating an integer programming formulation to be addressed to achieve a good tradeoff between test cost and prediction performance.
In this experiment, we set an MSE threshold to each specification as listed in TABLE \ref{setup-list}.
By solving Problem (\ref{0-1}), we obtain the optimal module selection represented by $\mathbf{x}^*$, in which $x_{1,2}^*=1$, $x_{1,3}^*=1$, $x_{2,3}^*=1$, and $x_{m,n}^*=0$ for other cases. 
This indicates that three test modules, 1-2, 1-3, and 2-3, are selected in our approach. \par
\begin{figure}[t]
   \centering
   \includegraphics[width=0.9\linewidth]{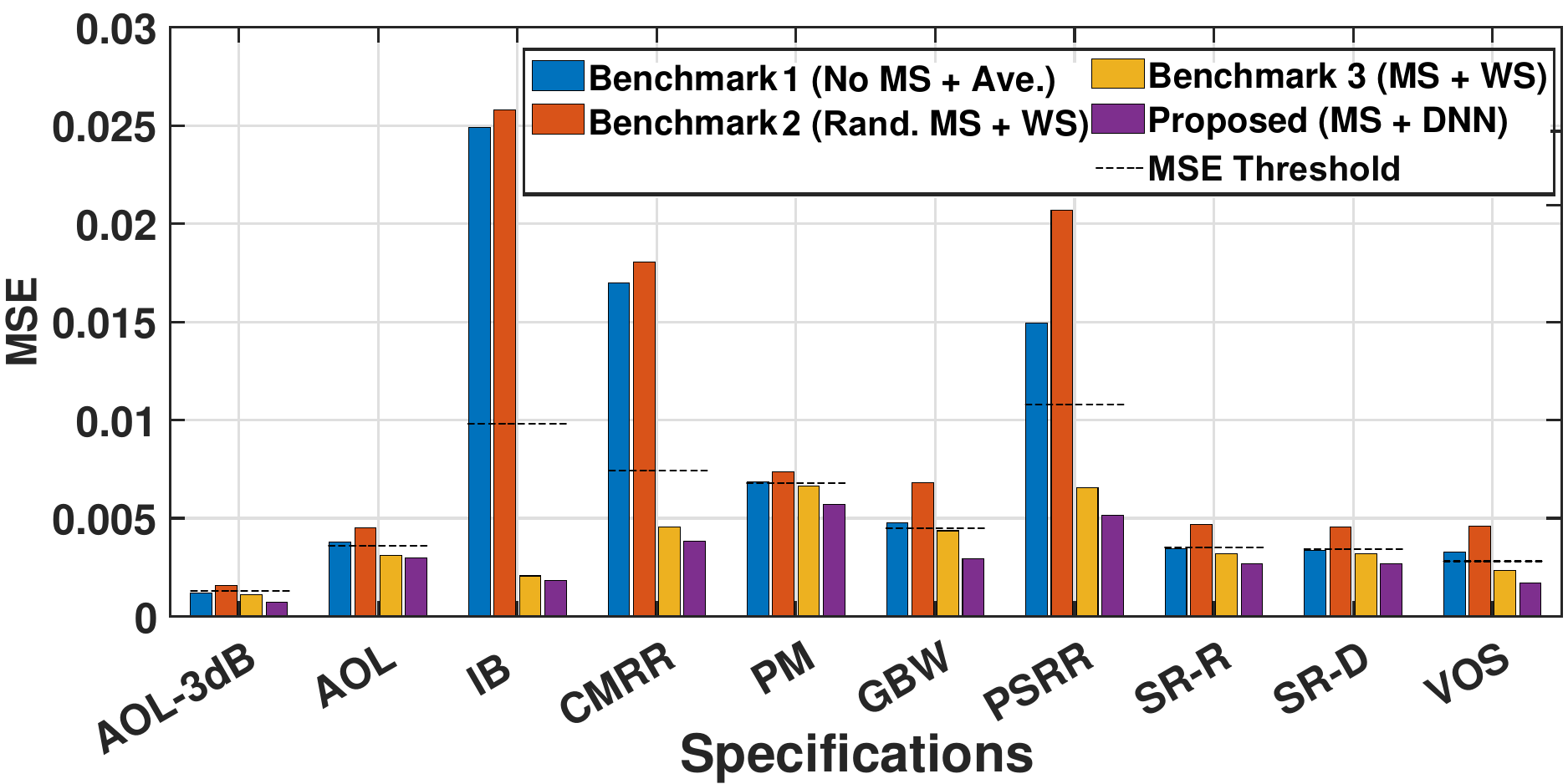}
   \caption{Comparing the MSE of every specification by proposed method and three benchmarks.}
   \label{mse-bar4}
   \vspace{-\baselineskip}
\end{figure}
In what follows, we compare the proposed scheme with three benchmark algorithms in Fig. \ref{mse-bar4}.
Benchmark 1 averages the predictions of all eight test modules, and Benchmark 2 makes a weighted sum of the predictions of three randomly selected test modules. 
Different from Benchmark 1 and Benchmark 2 that do not perform module selection, Benchmark 3 uses the same three modules as $\mathbf{x}^*$ indicated but derives the final specification estimation by weighted sums of the three modules' predictions. 
The MSE of each specification by Benchmark 1 and Benchmark 2 are higher than the MSE thresholds, indicating that these two baseline strategies cannot meet the practical performance requirement. 
Benchmark 1 and Benchmark 2, both without module selection, have worse prediction performance than those with module selection.
The main reason is that the modules with high MSE are employed without distinction by Benchmark 1 and Benchmark 2, where the MSE of the combined predictions will grow once these modules are selected.
Especially for IB, CMRR, and PSRR, their lower MSE only exist in a few modules but the gaps between the maximum and minimum MSE in different modules are noticeable.
If the modules with low MSE are not selected in a targeted manner, it will be challenging to guarantee test accuracy for these specifications.
Though both Benchmark 2 and Benchmark 3 adopt the weighted sum to combine the predictions, the MSE in Benchmark 3 has an apparent reduction compared with Benchmark 2, which validates the proposed module selection method.
Last but not least, the MSE of combining the selected predictions by the DNN is always lower than the weighted sum for each specification because the DNN performs nonlinear calibration on the predictions of the selected modules. \par

\begin{figure*}[h]
   \vspace{-\baselineskip}
    \centering
    \includegraphics[width=0.9\linewidth]{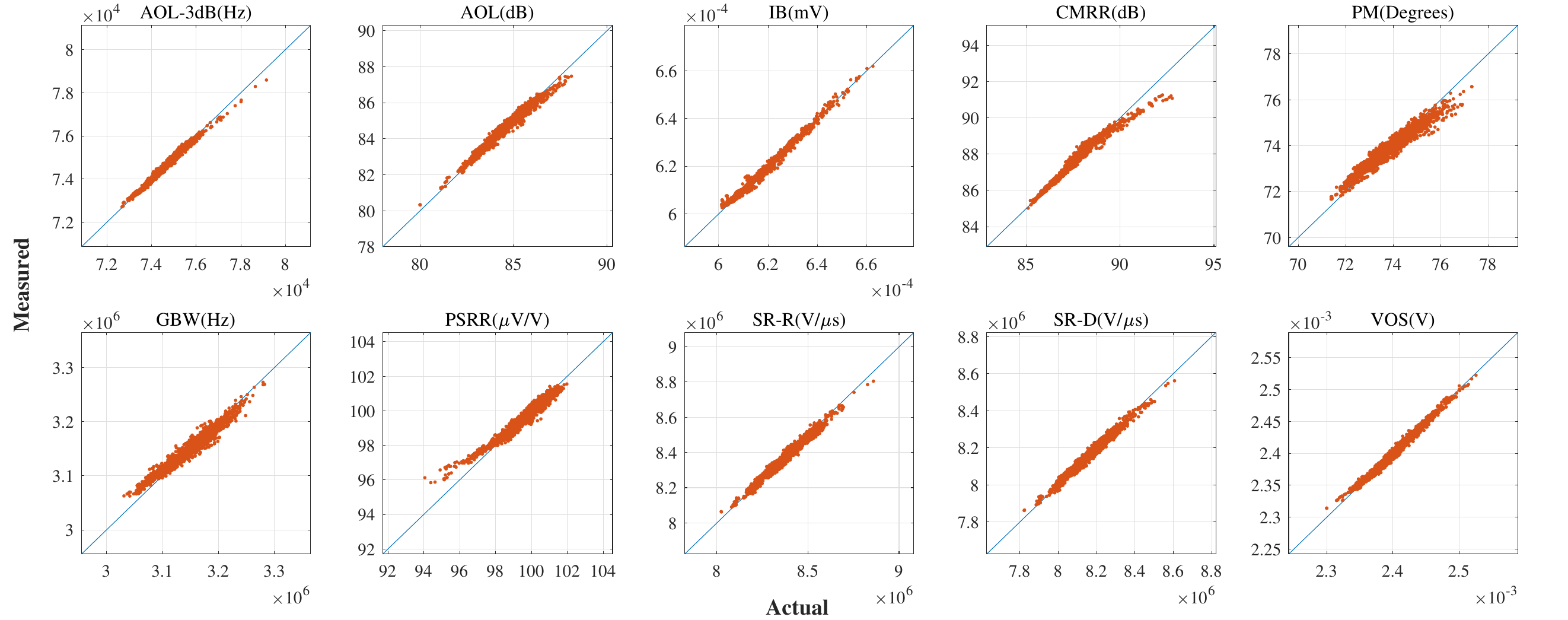}
    \vspace{-\baselineskip}
    \caption{Comparison between measured specifications and actual values.}
   \label{s1-10}
   \vspace{-1.5mm}
\end{figure*}

The two-dimensional data points that consist of the test and the actual values of every specification are depicted in Fig. \ref{s1-10}.
The distribution of the data points relative to the diagonal line in Fig. \ref{s1-10} and the MSE of proposed method could corroborate each other.
For example, the PM has the worst MSE, and the data points in Fig. \ref{s1-10}(e) are the most divergent relative to the diagonal line meanwhile;
conversely, the AOL-3dB has the lowest MSE, and the data points are nearly distributed along the diagonal in Fig. \ref{s1-10}(a).\par
\begin{table}[t]
   \vspace{-1.2\baselineskip}
   \renewcommand{\arraystretch}{1.2}
   \centering
   \caption{Fault Coverage}
   \label{fault coverage}
   \vspace{-1.5\baselineskip}
   \begin{tabular}[t]{cccccc}
   \midrule
   \textbf{Specification} & \textbf{AOL-3dB} & \textbf{AOL}     & \textbf{IB}     & \textbf{CMRR}  & \textbf{PM}\\ \hline
   \textbf{FC}  & 94.67\%   & 87.92\%   & 96.14\%   & 91.22\%  & 70.7\%      \\ \hline
   \textbf{Specification} & \textbf{GBW}     & \textbf{PSRR}    & \textbf{SR-R}   & \textbf{SR-D}  & \textbf{VOS}\\ \hline
   \textbf{FC}  & 75.1\%    & 82.73\%   & 94.44\%   & 97.11\%  & 94.53\%      \\ \hline
   \end{tabular}
   \vspace{-\baselineskip}
\end{table}

\begin{figure}[t]
   \centering
   \includegraphics[width=0.75\linewidth]{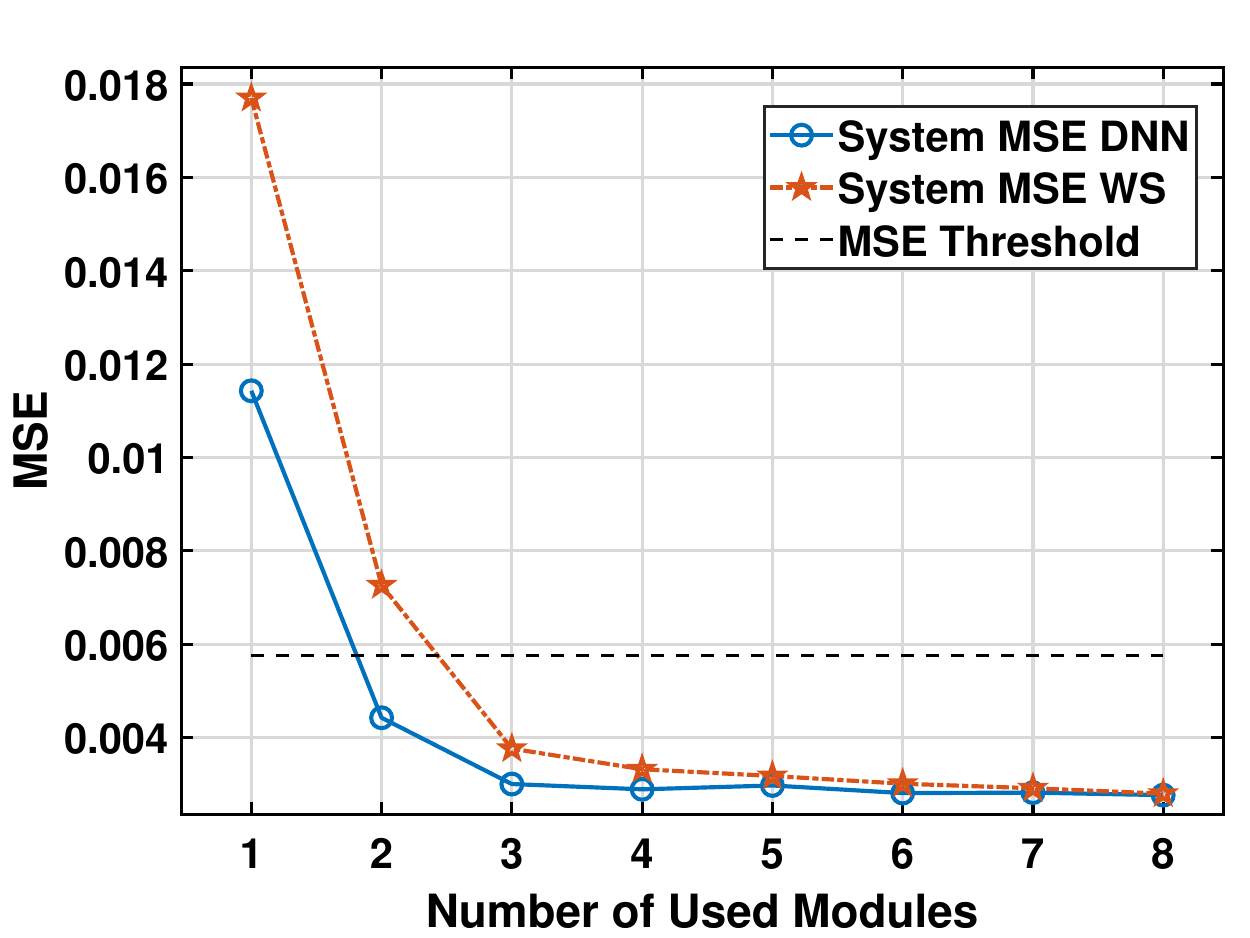}
   \vspace{-0.5\baselineskip}
   \caption{System MSE versus the number of used test modules.}
   \label{mse-r2}
   \vspace{-\baselineskip}
\end{figure}
From a different perspective, the fault coverage, i.e., the ratio of the number of detected faults to the total number of faults, could be used to evaluate the effectiveness of proposed method.
For a mass-produced IC, the fault-free range of each specification can be determined directly according to its datasheet.
For simulation in EDA tools, we take standard deviation $\sigma_\ell$ and mean $\mu_\ell$ for each specification to determine its fault boundary. 
Intervals $\left(-\infty, \mu_\ell+\sigma_\ell\right]$ and $\left[\mu_\ell-\sigma_\ell,+\infty\right)$ represent the fault-free ranges of specifications with one-sided boundary.
The predicted specifications outside their corresponding range are considered faults.
The FC by the proposed method is listed in the TABLE. \ref{fault coverage}.
Overall, the FC performance is commendable, the vast majority of test results that fall outside the fault-free range of the specification are correctly identified as faults.
The MSE of PM and GBW are lower than that of PSRR, nevertheless, PSRR has better FC than PM and GBW.\par
In the end, we average the MSE of all test specifications as the system MSE to evaluate the overall performance of the proposed framework.
The threshold in Fig. \ref{mse-r2} represents the averaging value of the MSE thresholds of all specifications.
On the whole, the system MSE reduces as the number of selected test modules increases.
While it contains a minimum value and reduces more slowly when it is closer to this minimum value, i.e., the improvement of the test accuracy becomes less noticeable when the test costs increase as the number of selected test modules increases.
The tradeoff between the system MSE and the test costs should be considered if trying to improve the test accuracy with the proposed testing framework.
Combining the predictions by a DNN is helpful for improving test accuracy. Thus, the system MSE by a DNN is always lower than that by the weighted sum, but the their gap decreases as the number of selected modules increases.

Particularly, when the system MSE satisfies the accuracy requirement, the required number of modules is three by the weighted sum, whereas the number is two by a DNN.
\par


\section{Conclusion}
This paper has proposed a low-cost testing framework for the specifications of analog ICs based on deep learning. 
For multiple test modules, we have employed different DNNs to establish the mapping from responses of the CUT in different test modules to specifications under test,
where different modules show varying capabilities for predicting the same specification.
We have modeled the test module selection issue as a 0-1 integer programming problem after specification-wise evaluating the MSE of predicted results of all test modules.
By solving the problem with an implicit enumeration algorithm, we have selected the minimum number of required modules for the test stage to reduce the test costs.
Finally, we have adopted a DNN again to combine the selected test results to further improve the test accuracy.
The simulation results have shown that the proposed testing framework can accurately test all specifications with the minimum test costs.


\bibliographystyle{ieeetr}

\begin{thebibliography}{8pt}
   \fontsize{8pt}{\baselineskip}
   \balance

   \bibitem{milor_optimal_1990}
   L.~Milor and A.~Sangiovanni-Vincentelli, ``Optimal test set design for analog circuits,'' in \emph{Proc. ICCAD}, 1990, pp. 294--297.

   \bibitem{huss_optimal_1991}
   S.~Huss and R.~Gyurcsik, ``Optimal ordering of analog integrated circuit tests to minimize test time,'' in \emph{Proc. DAC}, 1991, pp. 494--499.


   \bibitem{variyam_prediction_2002}
   P.~Variyam, S.~Cherubal, and A.~Chatterjee, ``Prediction of analog performance parameters using fast transient testing,'' \emph{IEEE Trans. Comput-Aided Des. Integr. Circuits Syst.}, vol.~21, no.~3, pp. 349--361, 2002.

   \bibitem{stratigopoulos_filter_2009}
   H.-G. Stratigopoulos, S.~Mir, E.~Acar, and S.~Ozev, ``Defect filter for alternate RF test,'' in \emph{IEEE Eur. Test Symp., ETS}, 2010, pp. 265--270.


   \bibitem{stratigopoulos_adaptive_2018}
   H.-G. Stratigopoulos and C.~Streitwieser, ``Adaptive test with test escape estimation for mixed-signal ICs,'' \emph{IEEE Trans. Comput-Aided Des. Integr. Circuits Syst.}, vol.~37, no.~10, pp. 2125--2138, 2018.

   \bibitem{badawi2021evaluation}
   H.~E. Badawi, F.~Aza{\"\i}s, S.~Bernard, M.~Comte, V.~Kerzerho, and F.~Lefevre, ``Evaluation of a two-tier adaptive indirect test flow for a front-end RF circuit,'' \emph{J. Electron. Test.-Theory Appl.}, vol.~37, pp. 225--242, 2021.

   \bibitem{deyati_dynamic_2020}
   S.~Deyati, B.~J. Muldrey, and A.~Chatterjee, ``Dynamic test stimulus adaptation for analog/RF circuits using booleanized models extracted from hardware,'' \emph{IEEE Trans. Comput-Aided Des. Integr. Circuits Syst.}, vol.~39, no.~10, pp. 2006--2019, 2020.

   \bibitem{banerjee_automatic_2015}
   A.~Banerjee and A.~Chatterjee, ``Automatic test stimulus generation for diagnosis of RF transceivers using model parameter estimation,'' \emph{IEEE Trans. Very Large Scale Integr. (VLSI) Syst.}, vol.~23, no.~12, pp. 3114--3118, 2015.

   \bibitem{barragan_use_2019}
   M.~J. Barragan, G.~Leger, F.~Cilici, E.~Lauga-Larroze, S.~Bourdel, and S.~Mir, ``On the use of causal feature selection in the context of machine-learning indirect test,'' in \emph{Proc. DATE}, 2019, pp. 276--279.

   \bibitem{leger_brownian_2016}
   G.~Leger and M.~J. Barragan, ``Brownian distance correlation-directed search: A fast feature selection technique for alternate test,'' \emph{Integr. VLSI J}, vol.~55, pp. 401--414, 2016.

   \bibitem{hou2022statistical}
   L.~Hou, Y.~Liu, W.~Xie, Z.~Dai, W.~Yang, and Y.~Zhao, ``Statistical neural network (SNN) for predicting signal-to-noise ratio (SNR) from static parameters and its validation in 16-bit, 125-msps analog-to-digital converters (ADCs),'' \emph{Rev. Sci. Instrum.}, vol.~93, no.~8, 2022.

   \bibitem{oneshot2016}
   M.~Andraud, H.-G. Stratigopoulos, E.~Simeu, ``One-shot non-intrusive calibration against process variations for analog/RF circuits,'' \emph{IEEE Trans. Circuits Syst. I, Reg. Papers1}, vol.~63, no.~11, 2022-2035, 2016.


   \bibitem{implicit}
   A. M. Geoffrion, ``Integer programming by implicit enumeration and Balas' method,'' in \emph{SIAM Rev.}, vol.~9, no.~2 pp. 178--190, 1967.
\end{thebibliography}

\end{document}